\documentclass[11pt]{article}

\usepackage{graphicx}
\title{RECENT RESULTS FROM SND DETECTOR AT VEPP-2M} 
\author{
M.N.~Achasov, V.M.~Aulchenko,  K.I.~Beloborodov, A.V.~Berdyugin,\\
A.V.~Bozhenok, A.G.~Bogdanchikov, A.D.~Bukin, D.A.~Bukin, S.V.~Burdin,\\
T.V.~Dimova, A.A.~Drozdetsky, V.P.~Druzhinin, V.B.~Golubev,\\ 
V.N.~Ivanchenko, P.M.~Ivanov, A.A.~Korol, S.V.~Koshuba, A.A.~Mamutkin, \\
E.V.~Pakhtusova, E.A.~Perevedentsev, E.E.~Pyata, A.A.~Salnikov, \\
S.I.~Serednyakov, V.V.~Shary, A.G.~Skripkin, Yu.M.~Shatunov, \\
V.A.~Sidorov, Z.K.~Silagadze, A.N.~Skrinsky, A.V.~Vasiljev \vspace{2mm}\\
{\em Budker Institute of Nuclear Physics, Novosibirsk 630090, Russia,}\\
{\em Novosibirsk State University, Novosibirsk 630090, Russia}\\
{ ~}\\
{presented by V.P.~Druzhinin}
}
\date{}
\begin{document}
\maketitle
\begin{abstract}
  The current status of experiments with SND detector at VEPP-2M
$e^+e^-$ collider in the energy range $2E_0=0.4-1.4$ GeV is
given. The new results of analysis of $\phi$ decay into $\pi^0\pi^0\gamma$, 
$\eta\pi^0\gamma$ are based on the full SND statistics
corresponding 20 million of $\phi$ decay. New measurement of
$\omega\to\pi^0\pi^0\gamma$ decay and a first observation of
$\rho\to\pi^0\pi^0\gamma$ are presented.  The accuracy of many
other rare decays of light vector mesons was improved.
In the energy range $2E_0=1.0\div1.4$ GeV the cross sections of the
processes $e^+e^-\to\omega\pi^0$ and $e^+e^-\to\pi^+\pi^-\pi^0$
were measured. The results of the fitting of data are discussed.
\end{abstract}
\section{Introduction}
VEPP-2M is the $e^+e^-$-collider \cite{VEPP2M}, operating
since 1974 in the energy range 2E=0.4--1.4 GeV
($\rho,\omega,\phi$-mesons region). Its maximum luminosity is about
$5\cdot10^{30}\,\mbox{cm}^{-2}\mbox{s}^{-1}$ at E=510 MeV. Two detectors
SND and CMD-2 carry out experiments at VEPP-2M now.

  SND was described in detail in \cite{SND}.
Its main part is the three layer spherical electromagnetic calorimeter 
consisting of 1620 NaI(Tl) crystals with a total mass of 3.6 tones. 
The solid angle coverage of the calorimeter is 90\% of $4\pi$ steradian. 
The energy resolution
for photons can be approximated as $\sigma_E(E)/E=4.2\%/E(\mbox{GeV})^{1/4}$,
angular resolution is about $1.5^{\circ}$. The angles of charged
particles are measured by two cylindrical drift chambers covering 98\%
of $4\pi$ solid angle.

  Since 1996 the SND detector collected $32~pb^{-1}$ of integrated luminosity in
three energy regions:
\begin{itemize}
\item {360--970 MeV}, $9\ pb^{-1}$ corresponding to $\sim7\times10^6$
produced $\rho$ mesons and $\sim4\times 10^6$ $\omega$  mesons; 
\item {980--1060 MeV}, $13\,pb^{-1}$ corresponding to $\sim2\times 10^7$
$\phi$ meson decays;
\item {1050--1380 MeV}, $9\ pb^{-1}$.
\end{itemize}
In this report we present results based on analysis of total statistics from last
two energy regions and $3.6~pb^{-1}$ from $\rho$, $\omega$ region.
\section {\boldmath Search for $\rho , \omega , \phi$ electric dipole 
radiative decays}
The decays of the vector mesons into a scalar and a photon are well known
for higher quarkonia, but there are very little data about such decays of light
mesons $\rho , \omega , \phi$. The scalar candidates for their decays are
$f_0(980)$,  $a_0(980)$ and  not  well established broad object  $\sigma(400-1200)$.\\

\begin{figure}[t]
\begin{minipage}[htb]{0.48\textwidth}
\includegraphics[width=0.95\textwidth]{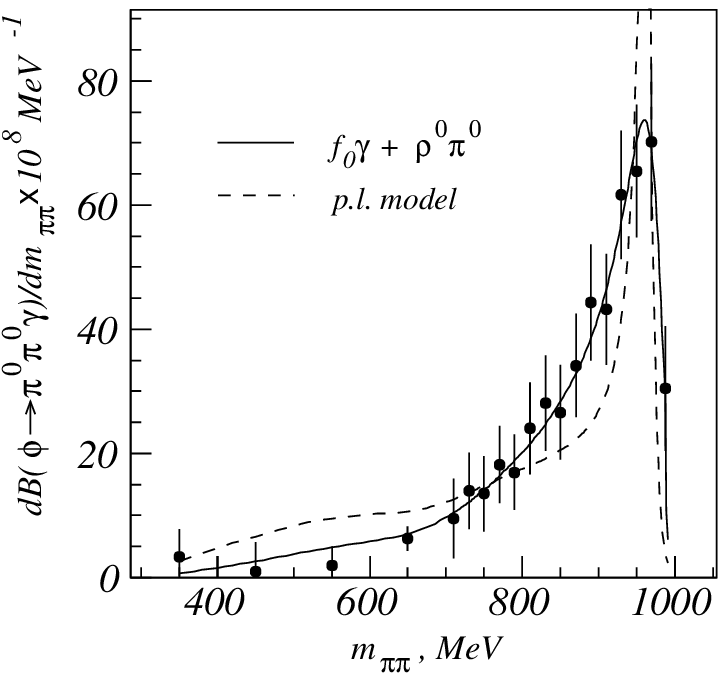}
\caption{The $\pi^0\pi^0$  mass in the decay $\phi\to\pi^0\pi^0\gamma$. }
\label{mpi0}
\end{minipage}
\hfill
\begin{minipage}[htb]{0.48\textwidth}
\includegraphics[width=0.95\textwidth]{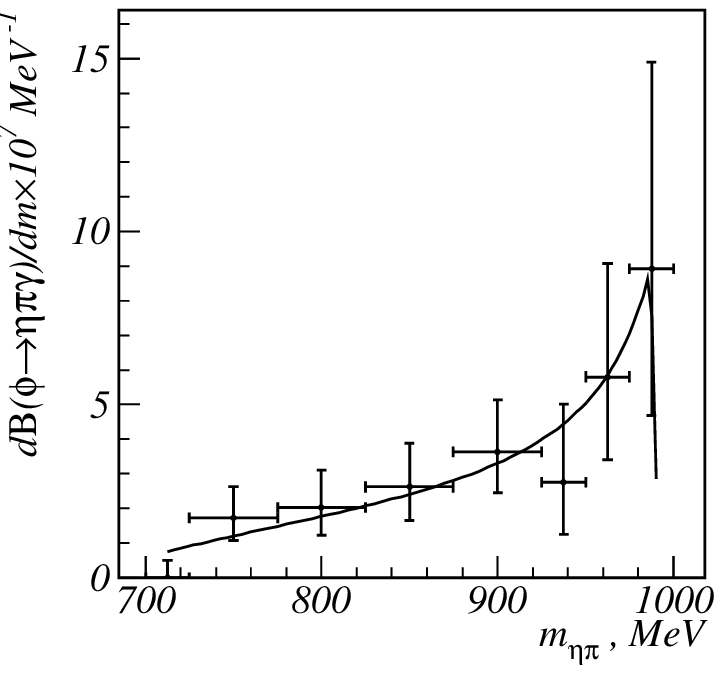}
\caption{The $\eta\pi^0$  mass  in the decay $\phi\to\eta\pi^0\gamma$.}
\label{metapi}
\end{minipage}
\end{figure}

{\bf \boldmath The decays  $\phi\to \pi^0\pi^0\gamma ,\eta\pi^0\gamma$.}
The first evidence of the electric dipole decays of $\phi$ meson 
was reported by SND detector in 1997 \cite{H97}. These decays were searched for
in the reactions:
\begin{equation}
e^+e^-\to\phi\to\pi^0\pi^0\gamma,  \label{pp0g} 
\end{equation}
\begin{equation}
e^+e^-\to\phi\to\eta\pi^0\gamma.   \label{etpg}
\end{equation}
On the base of the analysis of full SND data sample collected in the $\phi$ 
meson energy 
region the following branching ratios were obtained from 
the study of the reactions (\ref{pp0g}),  (\ref{etpg}) \cite{ppgf,epgf}:
\begin{equation}
B(\phi\to\pi^0\pi^0\gamma)=(1.22\pm 0.12)\cdot 10^{-4}, \label{bpp0g}
\end{equation}
\begin{equation}
B(\phi\to\eta\pi^0\gamma)=(0.88\pm 17)\cdot 10^{-4}. \label{betpg}
\end{equation}
Corresponding numbers of selected events were $419\pm31$ for the process 
(\ref{pp0g}) and $36\pm6$ for the process (\ref{etpg}). The angular 
distributions of these events were found to be in agreement with
expected for scalar intermediate $\pi^0\pi^0$ and $\eta\pi^0$ states. 
The $\pi^0\pi^0$ and $\eta\pi^0$ mass spectra
after background subtraction and applying the detection efficiency corrections
are shown in Figs.~\ref{mpi0},~\ref{metapi}. 
In spite of smaller recoil photon phase space
and $\sim E_\gamma$ dependence of an amplitude for the decay into scalar and 
photon both observed mass spectra demonstrate enhancements in higher 
mass regions. These enhancements can be explained by only resonant contribution 
of $f_0(980)$,  $a_0(980)$ mesons. The $\pi^0\pi^0$ mass spectrum was 
approximated by sum of contributions from $f_0(980)$ and $\sigma$ mesons with
a small addition of $\rho^0\pi^0$ mechanism calculated using VDM.
The $f_0(980)$ shape was described by Flatte \cite{flatte} type formula \cite{Pred} taking
into account the nearness of $K\bar{K}$ threshold. Results of the approximation
in the two models are shown in Fig.~\ref{mpi0}. In contrast to the point-like 
model of $\phi\to f_0\gamma$ transition which can not give satisfactory 
description of the data ($P(\chi^2)=28/14$), the model with the intermediate 
kaon loop \cite{Pred} well reproduce the shape of experimental
spectrum even without the additional contribution of $\sigma$ meson 
($P(\chi^2)=3/14$). The similar model was applied to describe the $\eta\pi^0$
mass spectrum in Fig.~\ref{metapi}. The fitting results demonstrate that 
$f_0\gamma$ and $a_0\gamma$ mechanisms dominate in the decays (\ref{pp0g}),
(\ref{etpg}). So, we can obtain from (\ref{bpp0g}) and (\ref{betpg}):
\begin{equation}
B(\phi\to f_0\gamma)=(3.5\pm 0.3^{+1.3}_{-0.5})\cdot 10^{-4} \label{bf00g}, 
\end{equation}
\begin{equation}
B(\phi\to a_0\gamma)=(0.88\pm 0.17) \cdot 10^{-4} \label{ba0g}.
\end{equation}
The result (\ref{bf00g}) was obtained assuming natural isotopic 
ratio $B(f_0\to\pi^+\pi^-)/B(f_0\to\pi^0\pi^0)=2$. 

It is hard to explain the relatively large values of $B(\phi\to f^0\gamma)$ 
and $B(\phi\to a^0\gamma)$ in the frame of a conventional two-quark description
of $f_0$ and $a_0$ structure (see discussion in the work \cite{kolya}). 
For example, the value of $B(\phi\to a^0\gamma)$ is close to 
$Br(\phi\to\eta^\prime\gamma)$. So, the isovector $a_0$ should contain strange
quarks like $\eta^\prime$! The possible solution is proposed by the four-quark MIT bag model
of $a_0$ and $f_0$ mesons which predictions are in a good agreement with our results 
\cite{Pred, kolya}. 
After observation of $\phi\to f_0\gamma ,a_0\gamma$ decays many works 
on $f_0$ and $a_0$ nature appeared \cite{altern}. All these models are 
different from a conventional
$q\bar{q}$ model and involve four-quark component either directly or as 
a result of strong S-wave meson-meson interaction.\\
\begin{figure}[t]
\begin{center}
\includegraphics[width=0.47\textwidth]{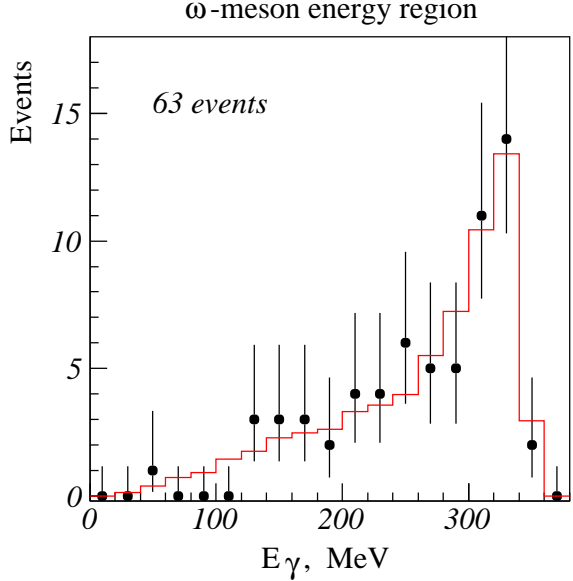}
\end{center}
\caption{The photon energy spectrum in the reaction $e^+e^-\to\pi^0\pi^0\gamma$
in the energy range near $\omega$ meson mass.}
\label{omsp}
\end{figure}

{\bf\boldmath {Search for the decay $\rho ,\omega\to \pi^0\pi^0\gamma$.}}
In VDM model these decays proceed  through the 
$\rho\to\omega\pi^0\to \pi^0\pi^0\gamma$ and
$\omega\to\rho\pi^0\to \pi^0\pi^0\gamma$ reactions with the relative 
probabilities  $\sim 10^{-5}$. With additional contribution $\sim 10^{-5}$ 
from pion chiral loops to $\rho\to \pi^0\pi^0\gamma$ decay, the branching 
ratios $B(\rho\to\pi^0\pi^0\gamma)=2.6\cdot 10^{-5}$  and
$B(\omega\to\pi^0\pi^0\gamma)=2.8\cdot 10^{-5}$ are predicted \cite{Bram}. The
only measurement of $\omega\to \pi^0\pi^0\gamma$ decay by GAMS \cite{GAMS} 
results value $(7.2\pm 2.5)\cdot 10^{-5}$, which is 
about three times larger than the theoretical expectation.

\begin{figure}[t]
\includegraphics[width=0.9\textwidth]{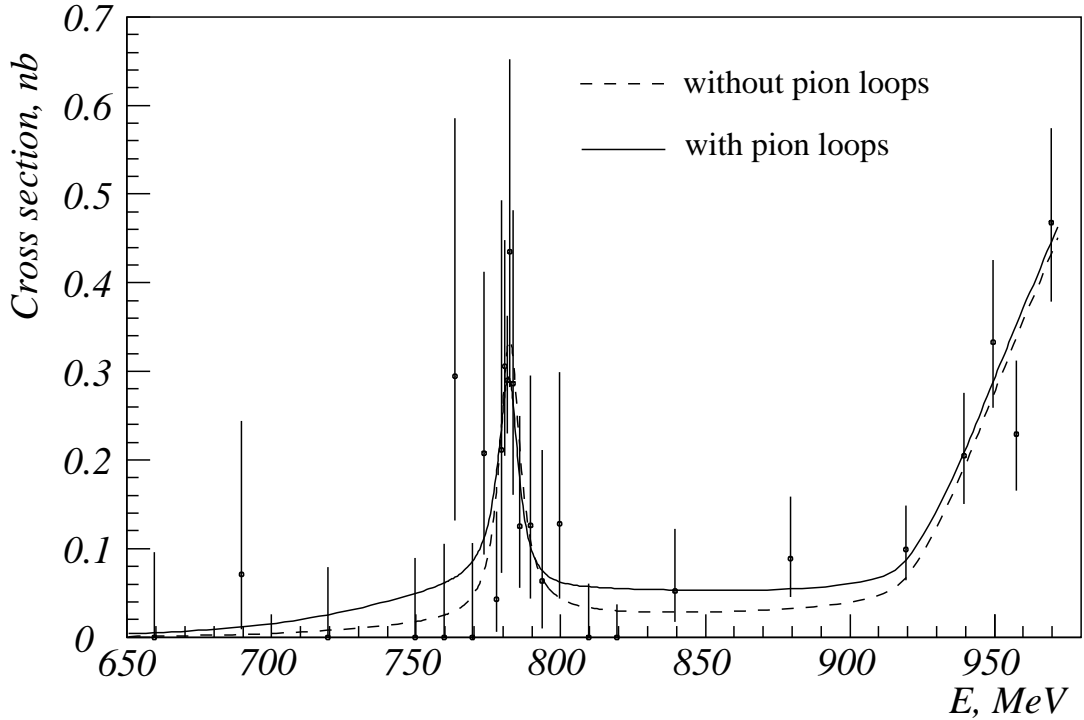}
\caption{The cross section of the $e^+e^-\to\pi^+\pi^-\gamma$ reaction and
the fitting curves for two models described in the text.}
\label{omer}
\end{figure}

About 150 pure events of the process $e^+e^-\to \pi^0\pi^0\gamma$ were 
selected in the energy region of  $\rho$ and $\omega$ resonances. 
The photon energy spectrum of events from the narrow energy range near 
$\omega$ are shown in Fig. \ref{omsp}. It is well described by
$\sim E_\gamma^3$ dependence expected for $S$-wave state of $\pi^0\pi^0$ pair.
But the problem is that in this energy range the $S$-wave contribution is 
dominant for all intermediate states including $\rho^0\pi^0$. So, we can not 
extract any information about $\omega\to \pi^0\pi^0\gamma$ decay mechanisms 
from the energy or angular distributions with our low statistics. The energy 
dependence of the $e^+e^-\to \pi^0\pi^0\gamma$ cross section is shown in 
Fig.~\ref{omer}.
The fit of the cross section included $\rho,\rho' \to\omega\pi^0$ transition 
and $\omega,\rho\to\pi^0\pi^0\gamma$ decays in the different models:
$\rho^0\pi^0$ and $S\gamma$. Here $S$ is $\sigma$ meson or S-wave 
$\pi^0\pi^0$ state in chiral pion loops mechanism. The strong
difference in the energy dependences of the phase spaces for 
$\rho\to\omega\pi^0$ and $\rho\to S\gamma$ mechanisms allows to distinguish 
the different models. The model without $\rho\to S\gamma$ contribution gives 
$P(\chi^2)=5\%$ and large value of 
$B(\omega\to\pi^0\pi^0\gamma )=(12.7\pm2.4)\times 10^{-5}$. Inclusion of the 
scalar mechanism to the fit improves $P(\chi^2)$ to 24\%. The resulting 
$\rho\to S\gamma$ amplitude was found to be 2.5$\sigma$ above zero. 

The branching ratios obtained from fitting of the cross section
are the following \cite{Vdruz}:
$$B(\omega\to\pi^0\pi^0\gamma)=(7.8\pm 3.3)\cdot 10^{-5} $$
$$B(\rho\to\pi^0\pi^0\gamma)=(4.8^{+3.4}_{-1.8})\cdot 10^{-5} $$
So, we have confirmed the value $B(\omega\to\pi^0\pi^0\gamma)$, obtained by GAMS. The decay
$\rho \to \pi^0\pi^0\gamma$ was observed for the first time. For both decays,
the measured values exceed the VDM predictions.

\subsection{Magnetic dipole radiative decays}
The magnetic dipole radiative decays $V\to P\gamma$ are traditional objects of
the study in the light meson spectroscopy. Only two
among the seven decays of this type, $\phi\to\eta\gamma$
and $\omega\to\pi^0\gamma$, are measured with relatively high accuracy.
The decay $\phi\to\eta^\prime\gamma$ was observed by CMD-2 not long ago, 
in 1997 \cite{cmd03}.

\begin{figure}[t]
\begin{minipage}[htb]{0.48\textwidth}
\includegraphics[width=0.95\textwidth]{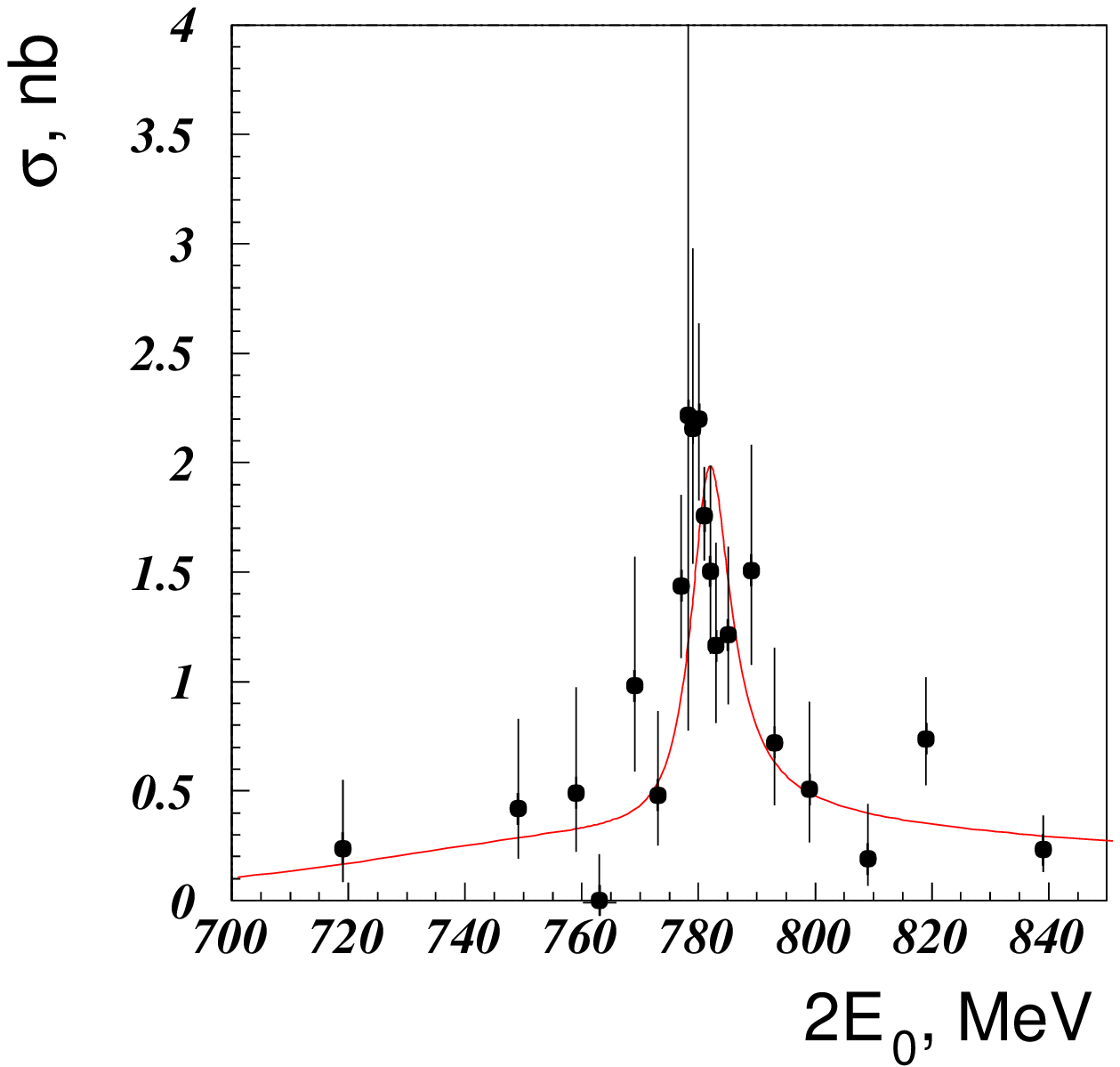}
\caption{The cross section of the reaction $e^+e^-\to \eta\gamma$ in
$\rho$ and $\omega$ energy region.}
\label{etagam1}
\end{minipage}
\hfill
\begin{minipage}[htb]{0.48\textwidth}
\includegraphics[width=0.95\textwidth]{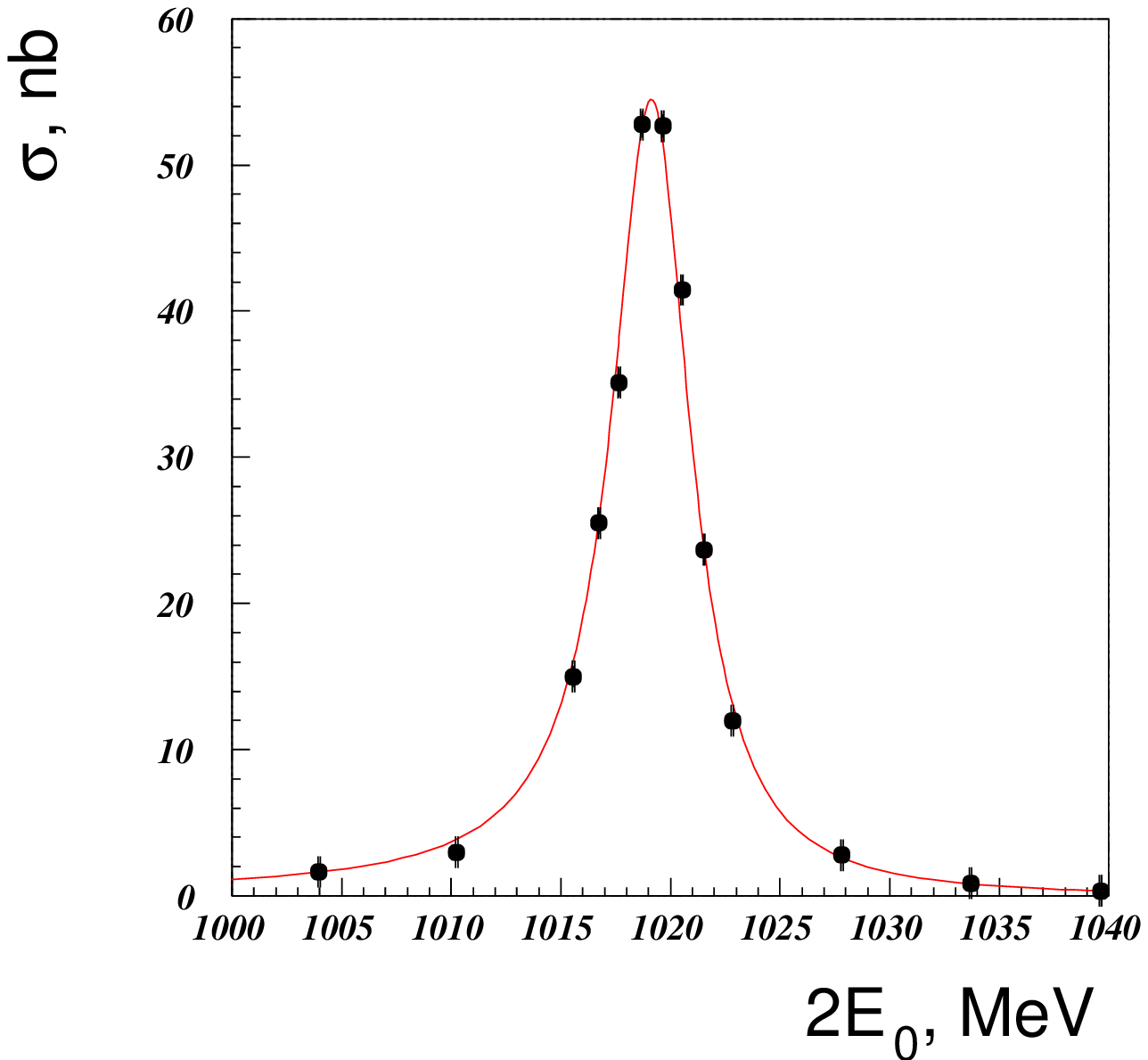}
\caption{The cross section of the reaction $e^+e^-\to \eta\gamma$ in
$\phi$ meson energy region.}
\label{etagam2}
\end{minipage}
\end{figure}

{\bf\boldmath $\rho,\;\omega,\;\phi\to\eta\gamma$ decays.} 
The reaction $e^+e^-\to 7\gamma$ is
free of any physical background and the best channel for study of 
$\rho, \omega\to\eta\gamma$ decays. The cross section of
the reaction $e^+e^-\to \eta\gamma$ measured in 7 photon final state is shown
in Figs. \ref{etagam1}, \ref{etagam2}. 
The results of fitting of cross section by
a sum of the contributions of $\rho$, $\omega$, and $\phi$ mesons are listed in
the following table \cite{berd}:
\begin{center}
\begin{tabular}{|l|c|c|} \hline
&SND ($7\gamma$ final state) & PDG98 \\ \hline
$\rho\to\eta\gamma$   & $(2.73\pm 0.31\pm 0.15)\times 10^{-4}$   & $(2.4\pm 0.9)\times 10^{-4}$ \\ 
$\omega\to\eta\gamma$ & $(4.62\pm 0.71\pm 0.18)\times 10^{-4}$   & $(6.5\pm 1.0)\times 10^{-4}$ \\ 
$\phi\to\eta\gamma$   & $(1.353\pm 0.011\pm 0.052)\times 10^{-2}$&$(1.26\pm 0.06)\times 10^{-2}$ \\
\hline
\end{tabular}
\end{center}
All three results have accuracies comparable or better than world average ones.
The experimental ratio of the partial widths
$\Gamma_{\omega\eta\gamma}:\Gamma_{\rho\eta\gamma}:\Gamma_{\phi\eta\gamma}=1:(15.4\pm2.6):(10.6\pm2.2)$
is in agreement with a prediction of the simple quark model:
$1:12:8$. 

The probability of the decay $\phi\to\eta\gamma$ was measured by SND
in two other decay modes of $\eta$ meson with following results:
$(1.259\pm0.030\pm0.059)\%$ for $\eta\to \pi^+\pi^-\pi^0$ \cite{eta3pi} and
$(1.338\pm0.012\pm0.052)\%$ for $\eta\to\gamma\gamma$ \cite{eta2gam}. 
Combining the 
results for three different modes we can obtain the SND average
$$BR(\phi\to\eta\gamma)=(1.310\pm0.045)\%,$$ the most precise
measurement of this value.

\begin{figure}[t]
\includegraphics[width=0.95\textwidth]{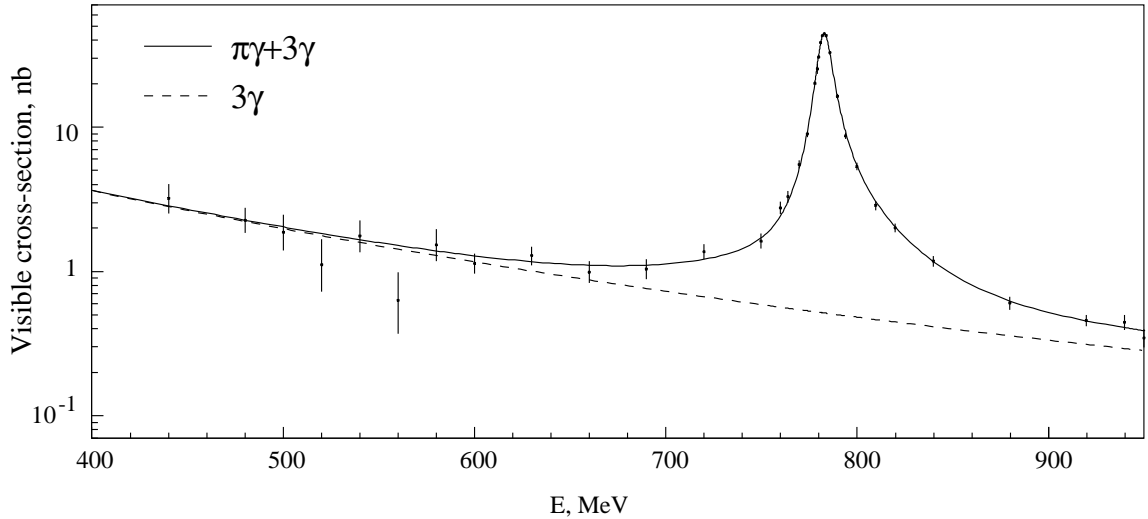}
\caption{The cross section of 3 photon events selected as candidates for
$e^+e^-\to\pi^0\gamma$ reaction.}
\label{pig}
\end{figure}

{\bf\boldmath $\rho,\;\omega\to\pi^0\gamma$ decays.} The cross section of 3 photon events 
selected as candidates for $e^+e^-\to\pi^0\gamma$ reaction is presented 
in Fig.~\ref{pig}.
The cross section was fitted by a sum of the contributions of
$\omega\to\pi^0\gamma$ and $\rho\to\pi^0\gamma$ decays and the background from
the process of $e^+e^-$ annihilation to three photon.  The preliminary results
of the fit together with
corresponding PDG values \cite{pdg} and SND result for $\phi\to\pi^0\gamma$ 
decay \cite{eta2gam}
are listed in following table:
\begin{center}
\begin{tabular}{|l|c|c|} \hline
&SND & PDG-1998 \\ \hline
$\rho\to\pi^0\gamma$  &$(4.3\pm2.2\pm0.4)\times 10^{-4}$ &$(6.8\pm 1.7)\times 10^{-4}$ \\
$\rho\to\pi^\pm\gamma$&                                            &$(4.5\pm 0.5)\times 10^{-4}$ \\
$\omega\to\pi^0\gamma$&$(8.5\pm0.2\pm 0.4)\times 10^{-2}$ &$(8.5\pm 0.5)\times 10^{-2}$ \\
$\phi\to\pi^0\gamma$  &$(1.23\pm 0.04\pm 0.09)\times 10^{-3}$     &$(1.31\pm 0.13)\times 10^{-3}$ \\ \hline
\end{tabular}
\end{center}
The $\rho, \omega\to\pi^0\gamma$ branching ratios are in a good agreement 
with both PDG values and a prediction of a simple quark model for 
$\rho\to\pi^0\gamma$ decay $\approx 5\times10^{-4}$ calculated
from $\omega\to\pi^0\gamma$ branching ratio. The obtained accuracies are comparable with table ones.
These results are based on a part of available statistics. For full data
sample we expect about two-fold
improvement of statistical accuracy of $\rho\to\pi^0\gamma$ 
branching ratio. We also hope
that combined analysis of data from $\phi$ and $\rho,\omega$ energy regions 
could reduce the systematic error of $\phi\to\pi^0\gamma$ branching ratio 
caused by the model dependence of
$\phi-\omega$ interference description.

\section{Rare $\phi$ decays}
\begin{figure}[p]
\begin{center}
\includegraphics[width=0.8\textwidth]{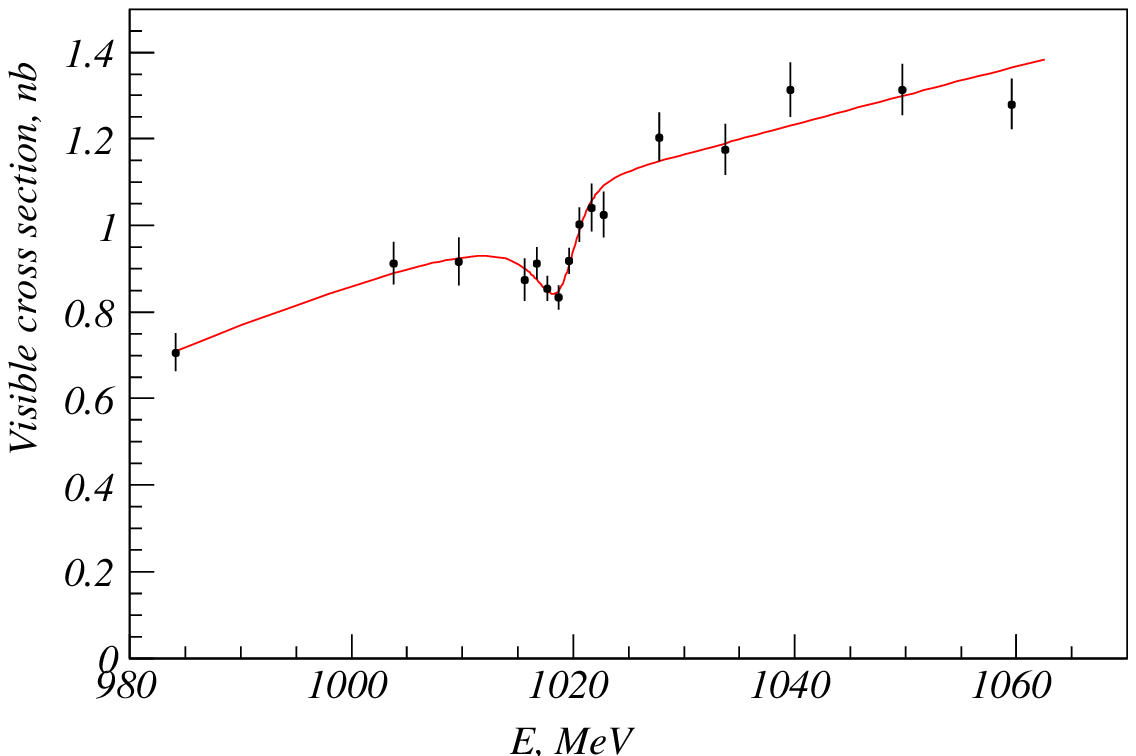}
\end{center}
\caption{The visible cross section of
$e^+e^-\to\omega\pi\to\pi^+\pi^-\pi^0\pi^0$ reaction near the $\phi$ peak.}
\label{ompn}
\vfill
\begin{center}
\includegraphics[width=0.65\textwidth]{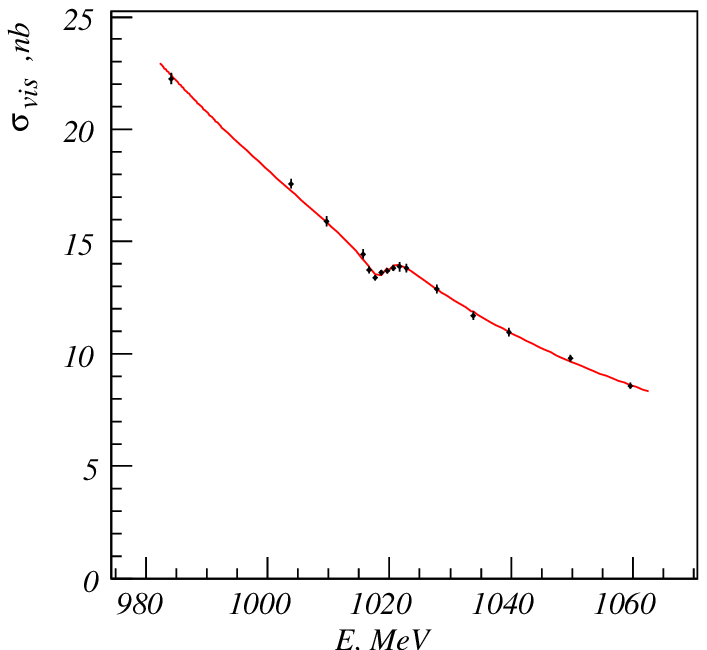}
\end{center}
\caption{The visible cross section of
$e^+e^-\to\pi^+\pi^-$ reaction near the $\phi$ peak.}
\label{pipi}
\end{figure}
{\bf\boldmath OZI and G-parity suppressed $\phi$ decays.}
The decays $\phi\to\pi^+\pi^-$, $\phi\to\omega\pi^0$ and $\phi\to\pi^+\pi^-\pi^0\pi^0$
were observed at VEPP-2M by detectors OLYA \cite{OLYA}, SND \cite{omp96} and
CMD-2 \cite{phi4p}. Here we will discuss the SND measurements of
$\phi\to\pi^+\pi^-$, $\phi\to\omega\pi^0$ decays. These double suppressed 
by QZI rule and G-parity decays can be seen as interference patterns in 
the energy dependenceof the 
cross sections of $e^+e^-\to\omega\pi$ and $e^+e^-\to\pi^+\pi^-$ processes. 
The Born cross section with the interference term 
can be written as follows:
$$\sigma(E) =\sigma_0(E)\cdot
\biggl |1-Z \frac{m_\phi\Gamma_\phi}{D_\phi(E)}\biggr |^2,$$
where $\sigma_0(E)$ is nonresonant cross section, $Z$ is complex
interference amplitude, $D_\phi(E)$ is $\phi$ meson inverse propagator.
One can extract from experimental data both real and imaginary parts
of the decay amplitude. The corresponding decay branching ratio is
proportional to $|Z|^2$ and $\sigma_0(m_\phi)$.
The simplest and most natural mechanism for
G-parity breaking is a single-photon transition $\phi-\gamma-\rho$ which
contributes only to real part of the interference amplitude:
$Re(Z)_{\gamma}=3B(\phi\to e^+e^-)/\alpha=0.123$. Other mechanisms 
are sensitive to the nature of $\rho-\omega-\phi$ mixing.

The cross-sections of selected events of $e^+e^-\to\omega\pi$ and
$e^+e^-\to\pi^+\pi^-$ processes for 1998 data set are shown in 
Figs.~\ref{ompn},~\ref{pipi}.
The interference patterns around $\phi$ meson mass are clearly seen in
both reactions.
The measured interference parameters and corresponding branching
ratios are listed in the following table \cite{omp98, pipi}:
\begin{center}
\begin{tabular}{|l|c|c|c|} \hline
 &Re(Z)&Im(Z)&$BR\times10^5$\\ \hline
$\phi\to\omega\pi^0$&$0.108\pm0.16$ &$-0.125\pm0.020$&$5.2^{+1.3}_{-1.1}$\\
$\phi\to\pi^+\pi^-$ &$0.061\pm0.006$&$-0.041\pm0.007$&$7.1\pm1.4$\\
\hline
\end{tabular}
\end{center}
The VDM model and standard $\omega -\phi$-mixing give considerably larger values of
branching ratios: $BR(\phi\to\omega\pi^0)=(8\div9)\times10^{-5}$ and
$BR(\phi\to\pi^+\pi^-)=34\times10^{-5}$.
The reasons of the discrepancy between the experiment and these predictions
are too low value of Re(Z), measured in both decays.
A possible explanation are considered in \cite{kozh, oller} and 
could be a nonstandard $\omega -\phi$-mixing and 
direct decays  $\phi\to\pi\pi$,  $\phi\to\omega\pi^0$.

The decay $\phi\to\omega\pi^0$ was observed by SND for the first time. The
measured $\phi\to\pi^+\pi^-$ branching ratio agrees with PDG value \cite{pdg}:
$(8^{+5}_{-4})\cdot 10^{-5}$ but is in contradiction
with preliminary CMD-2 result $(18\pm 3)\cdot 10^{-5}$ \cite{cmd04}.\\

{\bf\boldmath $\phi$ meson leptonic branching ratios.}
The usual and most precise method of the determination of $\phi$ meson
leptonic branching ratio is an extraction of $B(\phi\to e^+e^-)$
from the value of the $\phi$ production cross section in $e^+e^-$ collisions.
This cross section is measured as a sum of all $\phi$ decay modes:
$\phi \to K^+K^-, \: K_S K_L, \: 3\pi$, etc. The list of
the branching ratios of the main $\phi$ decay modes measured by SND \cite{mach}
is presented in the following table:
\begin{center}
\begin{tabular}[t]{|l|c|c|}\hline
  & SND & PDG98 \\ \hline
  $B(\phi \to K^+K^-), \%$ & $47.4 \pm 1.6 $ & $49.1 \pm 0.8$ \\
  $B(\phi \to K_SK_L), \%$ & $35.4 \pm 1.1$ & $34.1 \pm 0.6$  \\
  $B(\phi \to 3\pi), \%$   & $15.9 \pm 0.7$ & $15.5 \pm 0.7$  \\
  \hline
  $B(\phi \to e^+e^-) \times 10^{4}$ & $2.94 \pm 0.14$ & $2.99 \pm 0.08$ \\
  \hline
  \end{tabular}
  \end{center}
The last line of the table shows $B(\phi\to e^+e^-)$ value obtained by SND.

\begin{figure}[t]
\includegraphics[width=0.8\textwidth]{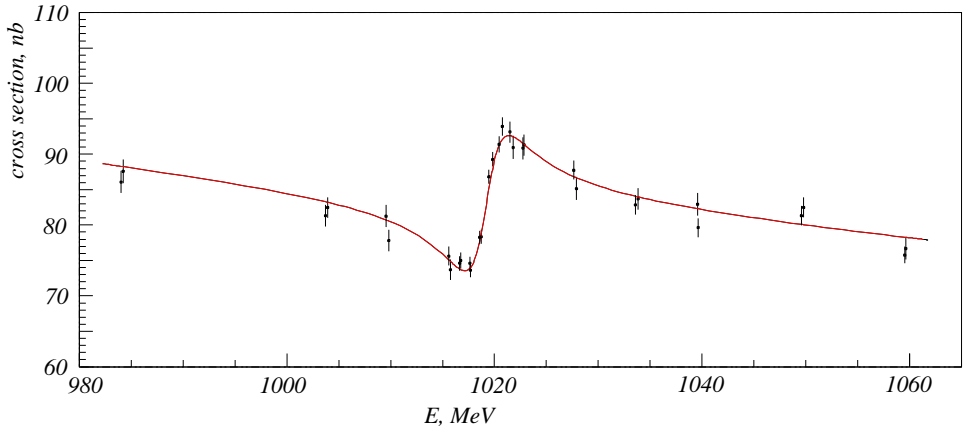}
\caption{The visible cross section of
$e^+e^-\to\mu^+\mu^-$ reaction near the $\phi$ peak.}
\label{mumu}
\end{figure}
Another method of the determination of the leptonic width is measurement of 
the amplitude of interference pattern
in the cross section of $e^+e^-\to\mu^+\mu^-$. This amplitude is
equal to $\sim12\%$ and proportional to $\sqrt{B(\phi\to e^+e^-)B(\phi\to \mu^+\mu^-)}$.
Up to now an accuracy of this method was limited by experimental statistics.
The Fig.~\ref{mumu} demonstrates the $e^+e^-\to\mu^+\mu^-$ cross section
in $\phi$ meson energy region measured by SND detector.
From the fit of experimental cross section we obtain the following
value of $\phi$ meson leptonic branching ratio \cite{mumu}:
$$\sqrt{B(\phi\to e^+e^-)B(\phi\to \mu^+\mu^-)}=(2.93\pm 0.10\pm 0.06)\cdot 10^{-4}$$
which is in a good agreement with $B(\phi\to e^+e^-)$ value and has
comparable accuracy. Using table value of $B(\phi\to e^+e^-)$
we can obtain the probability of $\phi\to\mu^+\mu^-$ decay \cite{mumu}:
$$B(\phi\to\mu^+\mu^-)=(2.87\pm0.20\pm 0.14)\cdot 10^{-4}$$
Our result is the most precise measurement of $B(\phi\to\mu^+\mu^-)$. 
\section{$e^+e^-$ annihilation into hadrons.}
The process of $e^+e^-$ annihilation into hadrons in the 1--2 GeV energy region
is an important source of information about excited
states of light vector mesons $\rho$, $\omega$ and $\phi$.
The current PDG status \cite{pdg} of these states based mainly on the analysis
of $e^+e^-$ annihilation cross sections and $\tau$ lepton hadronic
decays by A.B.Clegg and A.Donnachie \cite{clegg} are shown in the following table:
\begin{center}
\begin{tabular}[t]{|l|c|c|c|c|}\hline
 &$\rho^\prime$ & $\rho^{\prime\prime}$ & $\omega^\prime$ &$\omega^{\prime\prime}$\\ \hline
 Mass, MeV  &$1465\pm25$& $1700 \pm 20$  &$1419\pm31$& $1649 \pm 24$ \\
 Width, MeV &$310\pm60$& $240 \pm 60$  &$174\pm60$& $220 \pm 35$ \\
 \hline
 \end{tabular}
 \end{center}
The key channels for $\rho^\prime$ and $\omega^\prime$ states are 
$e^+e^-\to\pi^+\pi^-,\: \omega\pi,\: \pi^+\pi^-\pi^0$ reactions. 
Recently new data in this energy region became available from
SND \cite{m3pi,momp}, CMD-2 \cite{cmd4p}, CLEO \cite{cleo4p, cleo2p}, 
ALEPH\cite{aleph} experiments. We present the results of SND 
measurements of $e^+e^-\to \omega\pi$ \cite{momp} and 
$e^+e^- \to \pi^+\pi^-\pi^0$ \cite{m3pi} cross 
sections at the energy up to 1.4 GeV.

\begin{figure}[t]
\includegraphics[width=0.95\textwidth]{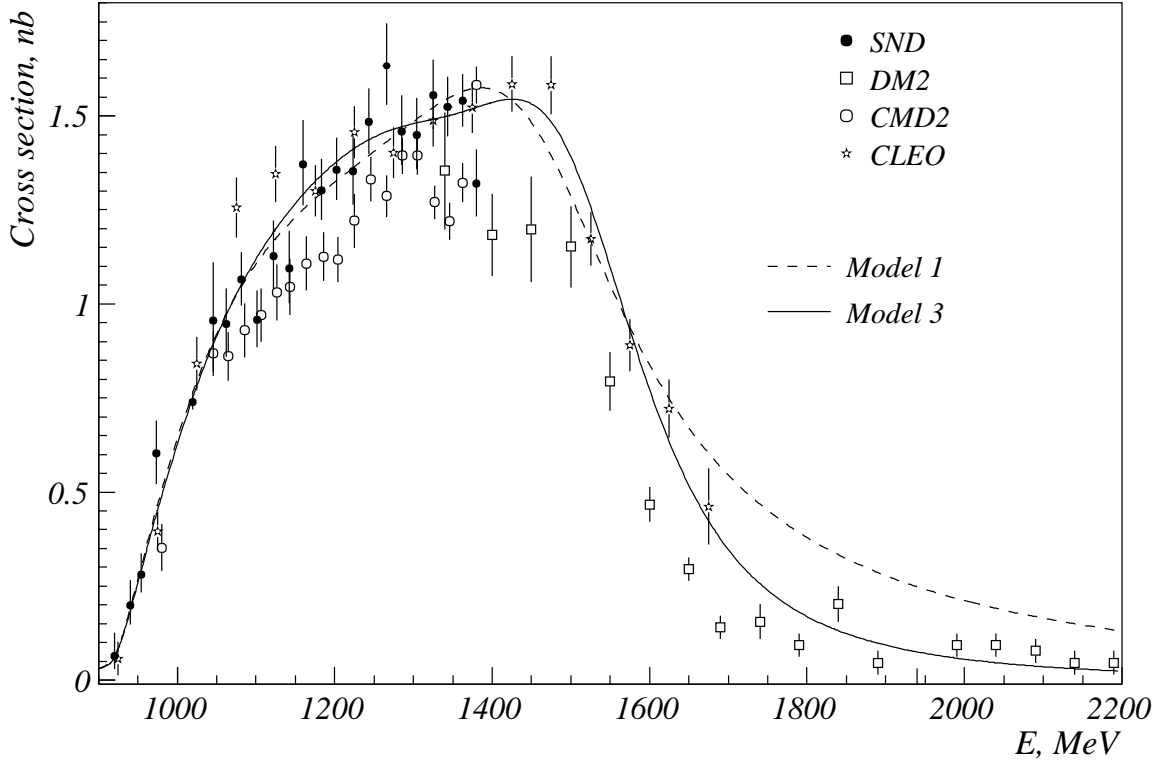}
\caption{The cross section of the reaction
$e^+e^-\to\omega\pi^0\to\pi^0\pi^0\gamma$. The results
of the SND \cite{momp}, DM2 \cite{dm2-4pi}, CMD \cite{cmd4p},
CLEO \cite{cleo4p} experiments are shown. Curves are
results of fitting to the data in model 1 and 3 with $R=0$.}
\label{f3}
\end{figure}

{\bf\boldmath Process $e^+e^-\to \omega\pi\to \pi^0\pi^0\gamma$.}
The process $e^+e^-\to \omega\pi$ was studied in five photon 
$\pi^0\pi^0\gamma$
final state in which this intermediate state is dominant.
Measured cross section in comparison with the most precise
CMD-2~\cite{cmd4p}, CLEO~\cite{cleo4p}, and DM2~\cite{dm2-4pi}
measurements are shown in Fig.\ref{f3}.
The CLEO results are in good
agreement with ours while the CMD-2 measurements are about 10\%
lower, although
the difference observed is smaller than the 15\%
systematic error quoted in
\cite{cmd4p}. There is a significant difference between the results of
DM2 and CLEO. For the cross section fitting we used our data together
with the data from CLEO.
The energy dependence of the cross section was described by a sum of
contributions of $\rho(770)$ and its excitations $\rho^\prime$ and $\rho^{\prime\prime}$.
Two different approaches were considered to describe of $\rho^\prime$ and $\rho^{\prime\prime}$
shapes. One of them \cite{clegg} assumes constant total width of
excited states (Model 1). In another one \cite{Achasov-rho} energy dependent width
is used: $\Gamma_{\rho_i}\sim q^3/(1+(qR)^2)$, where $q$ is momentum of $\omega$ meson
in $\omega\pi$ final state, $R$ is parameter restricting fast growth
of the resonance width (Model 2 and 3). The fit parameters obtained in 3 models with $R$
ranged from 0 to 2 GeV$^{-1}$ are listed in following table:
\begin{center}
\begin{tabular}{|c|c|c|c|}
\hline
& Model 1& Model 2& Model 3 \\
\cline{2-4}
$m_{\rho^\prime}$, MeV&1460--1520& -- &$\equiv1400$\\
$\Gamma_{\rho^\prime}$, MeV&380--500& --&$\equiv500$\\
$m_{\rho^{\prime\prime}}$, MeV&--&1710--1580&1620--1550\\
$\Gamma_{\rho^{\prime\prime}}$, MeV&--&1040--490&580--350\\
$\chi^2/N_D$&(52--48)/35&(47--48)/35&(43--44)/34\\
\hline
\end{tabular}
\end{center}
Both models 1 and 2 consider only one excited $\rho$ state but
give very different results. An inclusion of the energy dependent width
in the model 2 leads to significant growth of resulting mass and
width of the exited state. Only in model 1 with $R=0$ the parameters
$\rho^\prime$ meson are compatible with their PDG values, but
this model yields a poorest $\chi^2$ value: $P(\chi^2)=3\%$.
The satisfactory description of the experimental data was obtained
in model 3 with two excited states. The mass and width of first one
were fixed to 1400 MeV and 500 MeV respectively. These parameters were
taken from CLEO analysis of
$\pi^+\pi^-$ channel\cite{cleo4p}. However the large amplitude of
$\rho^{\prime\prime}$ meson obtained in this case contradicts
the theoretical expectations \cite{Barnes,Isgur} which predict larger
contribution from the lowest excited state $\rho^\prime$.

\begin{figure}[t]
\begin{minipage}[htb]{0.48\textwidth}
\includegraphics[width=0.95\textwidth]{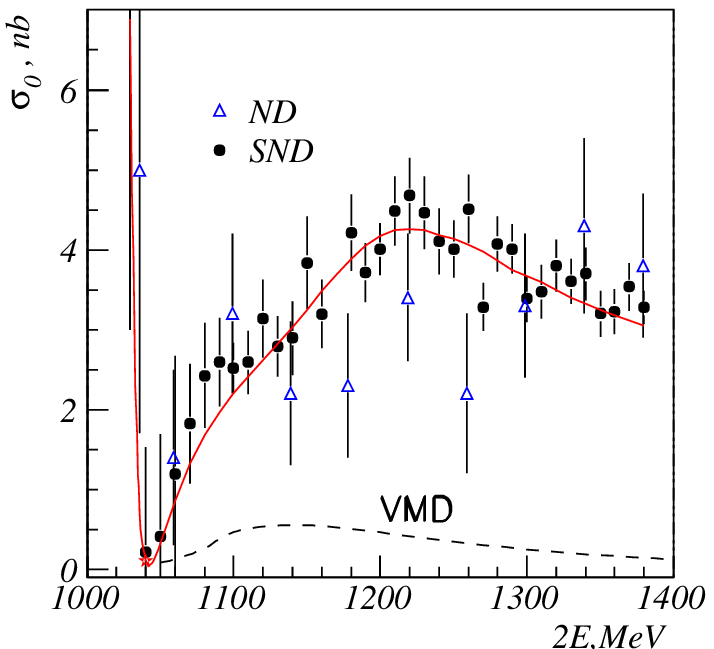}
\caption{The cross section of the reaction $e^+e^-\to \pi^+\pi^-\pi^0$.
The lower curve is a prediction of VDM with only
$\omega(782)$ and $\phi(1020)$ contribution.}
\label{cr3p}
\end{minipage}
\hfill
\begin{minipage}[htb]{0.48\textwidth}
\includegraphics[width=0.95\textwidth]{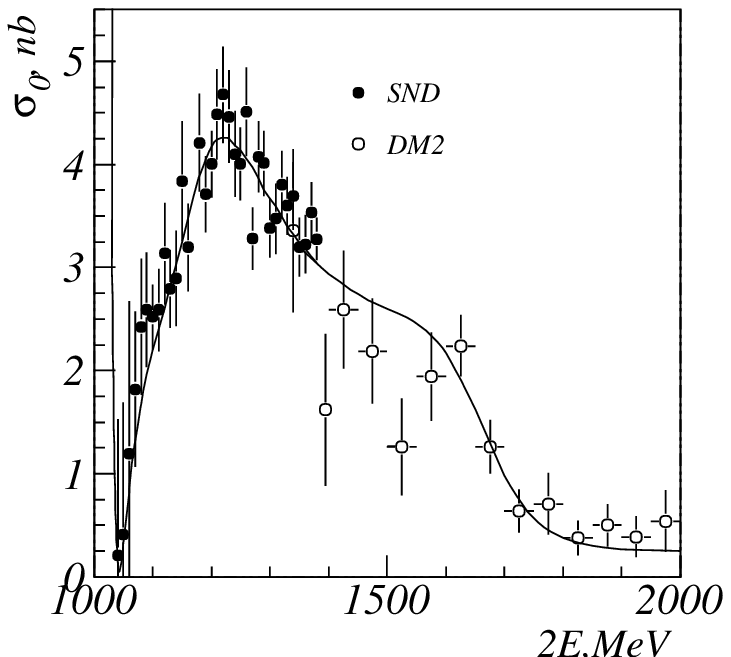}
\caption{The cross section of the reaction $e^+e^-\to \pi^+\pi^-\pi^0$.
The SND \cite{omprime} and DM2 \cite{Dm2} data are shown. 
The curve is a fit result.}
\label{cr3p1}
\end{minipage}
\end{figure}

{\bf\boldmath Process $e^+e^-\to \pi^+\pi^-\pi^0$.} The
result of SND measurements of $e^+e^-\to \pi^+\pi^-\pi^0$ 
is presented in Fig. \ref{cr3p}.
The measured cross section shows a broad maximum at $2E\simeq 1200$ MeV.
The SND and DM2 \cite{Dm2} data (Fig. \ref{cr3p1}) were fitted by a sum of 
$\phi$, $\omega$, $\omega^\prime$, $\omega^{\prime\prime}$ amplitudes. Similar
to $e^+e^-\to\omega\pi$ case the fit
gives $\omega^\prime$ parameters strongly dependent on the  model used.
For example, in the model with $\Gamma_{\omega^\prime}$=constant we obtained
$M_{\omega^\prime}=1170\div 1250$ MeV, $\Gamma_{\omega^\prime}=190\div 550$ MeV
\cite{omprime}, while
the model with strong width dependence on the energy  gives  $\omega^\prime$
parameters $M_{\omega^\prime}=1430\pm 100$ MeV, 
$\Gamma_{\omega^\prime}\sim 900$ MeV \cite{achopr}
which are close to the PDG  values.\\ 

The conclusions from the analysis of the processes $e^+e^-\to\omega\pi^0$
and $e^+e^-\to\pi^+\pi^-\pi^0$ are the following.
Fitting of the same experimental data by
models with fixed and energy-dependent total widths of the excited states
yields quite different parameters of these states.
This is caused by strong energy dependence
of the phase space for the main decay modes of $\rho^\prime$ and
$\omega^{\prime}$ mesons and this effect should be
taken into account in the fitting of experimental data.
To obtain more definite values of the parameters of $\rho$ and $\omega$
exited states new experimental data at higher energies 
$2E=1400\div 2000$ MeV are needed.
We hope that these data will be soon available from experiments
at VEPP-2000 $e^+e^-$ collider \cite{pdg2000} which construction is to be
started in 2000 in BINP, Novosibirsk. The two upgraded detectors SND and CMD-2 
will take data at VEPP-2000 with $1 fb^{-1}$ of integrated luminosity. 
The physical program is aimed to detailed study of $e^+e^-$ annihilation 
processes in the energy range $2E_0=1\div 2$ GeV.
\section{Acknowledgment}
The authors are grateful to N.N.Achasov for useful discussions.
The work is partially supported by RFBR (Grants No 00-15-96802, 00-02-17481,
99-02-16815, 99-02-16813) and STP ``Integration'' (Grant No 274).

\end{document}